\documentclass[12pt,twoside]{article}
\usepackage{graphicx}
\usepackage[english]{babel}
\usepackage[cp866nav]{inputenc}

\newcommand{\no}{\nonumber}
\newcommand{\non}{\nonumber \\}
\newcommand{\ve}[1]{{\bf #1}}
\newcommand{\be}{\begin{equation}}
\newcommand{\ee}{\end{equation}}
\newcommand{\bea}{\begin{eqnarray}}
\newcommand{\eea}{\end{eqnarray}}

\newcommand{\lb}{\left \{}

\newcommand{\rd}{\right .}

\newcommand{\vk}{\ve{k}}
\newcommand{\vl}{\ve{l}}
\newcommand{\vj}{\ve{j}}

\begin{document}


\title{Free Energy and Equation of State of Ising-like Magnet Near the Critical Point}

\maketitle
\author{M.P. Kozlovskii},
\author{I.V. Pylyuk},
\author{O.O. Prytula}

\begin{abstract}
The description of a three-dimensional Ising-like magnet in the
presence of an external field in the vicinity of the critical
point by the collective variables method is proposed. Using the
renormalization group transformations, the scaling region size
is defined as a function of temperature and field.
The obtained expressions for the free energy, equation of state and
susceptibility allow one to analyse their dependence on microscopic
parameters of the system. The critical exponents of the correlation
length and order parameter are calculated as well. The results agree qualitatively
with ones obtained within the framework of the parametric representation of
the equation of state and Monte-Carlo simulations. The calculations do
not involve any parametrization, phenomenological assumptions and
adjustable parameters. The approach can be extended to models with a
multicomponent order parameter.
\end{abstract}




\section{Introduction}
The description of the phase transitions in the three-dimensional ($3D$)
magnets, usually, is associated with the absence of exact solutions and with many approximate approaches
for obtaining different system characteristics \cite{palvi102}. Even the simplest Ising model, despite the
variety of works devoted to its study, is still of interest \cite{m05}.
Inclusion of the external factors affecting on the system behaviour complicates the problem.

Some of the more popular approaches, used for the study of critical phenomena,
are based on the series expansions of different characteristics
in small parameters, scaling theory, renormalization group (RG) transformations, etc.

Let us recall two kinds of the RG theory.
One of them is based on the perturbation theory (see, for example, \cite{bs97}).
The non-Gaussian distributions are used in another one \cite{btw02}.
An advantage of the non-perturbative methods is the possibility to avoid so-called
infrared (IR) problems, which exist in the pertubative approaches.
One of the most popular  non-perturbative approaches employs
the exact RG equation (see \cite{btw02,tw94,bb01}).
The application of the
non-perturbative methods is effective for investigating the Ising universal behaviour in
high energy physics. In \cite{t02}, the critical exponents for systems in the universality class
of the $3D$ Ising model are computed in the local potential approximation (LPA), that is,
within the framework of the Wegner-Houghton equation.
The similar calculations are performed for different IR cutoffs, which can be chosen freely \cite{lv04}.
In both papers \cite{t02} and \cite{lv04}, the effective potential contains the odd part, which allows one
to calculate the critical exponents related to the field dependence.

We propose the description of the critical behaviour of a $3D$ Ising-like magnet
in the presence of an external field by some alternative non-perturbative approach
called collective variables (CV) method
\cite{prfref5,novo89}. It exhibits some similar features  of the Wilson exact RG approach.
In particular, the Kadanoff hypothesis for the construction of the effective lattice, which is
commensurate with the size of long-range interactions, is used in both methods.
In contrast to other methods, the calculations carried out within the CV approach are performed on the
microscopic level without any phenomenological assumptions and adjustable parameters. Star\-ting from
the Hamiltonian, the CV method makes it possible to obtain the critical exponents as well as
the ther\-mo\-dy\-na\-mic and structural  system characteristics. In particular, we get
the free energy, equation of state and susceptibility as functions of temperature
and field.

Using this method, we have already obtained thermodynamic characteristics in the form
of a power series in the scaling variable in the regions of the weak and strong fields for
temperatures above and below $T_c$ ($T_c$ is the phase transition temperature in the absence
of an external field) \cite{kppfer05,kpprevb06}. In the mentioned papers as well as in \cite{mycmp05},
the size of the scaling region has been defined only by the temperature or by the field.
In additional to even powers of CV, the effective Hamiltonian has contained the odd part represented
by the linear and cubic terms, which are related to the presence of the field.
In contrast to the works \cite{kppfer05,kpprevb06}, the present calculations are carried
out without using power series in scaling variable. The size of the scaling region is a function of
both the temperature and field variables. Due to removing the cubic term (by the corresponding substitution
of variables), the odd part of the effective Hamiltonian is represented only by the linear term. The obtained
results are valid in the whole field-temperature plane.

\section{RG transformations}
The system Hamiltonian
\be
H=-\frac{1}{2}\sum_{\vl,\vj}\Phi(r_{\vl \vj})
\sigma_\vl\sigma_{\vj}-h\sum_{\vl}\sigma_{\vl}
\label{f1}
\ee
includes the interaction potential $\Phi(r_{\vl \vj})=A\exp{(-r_{\vl\vj}/b)}$, which is the function
of the distance between spins  $\sigma_{\vl}$ located at sites $\vl$ and $\vj$ of a
simple cubic lattice. The microscopic parameter $b$ is the radius of effective interaction, $A$
is a constant. The second term on the right-hand side of formula (\ref{f1}) is related to the
presence of an external magnetic field $h$.

The CV $\rho_{\vk}$ are introduced by means of the functional representation for
operators \cite{prfref5,novo89}
\be
\hat{\rho_\vk}=(\sqrt{N})^{-1}\sum_{\vl}{\sigma_{\vl}\exp(-i\vk\vl)}.
\label{f2}
\ee
Using the Jacobian of transition from the set of $N$ spin variables $\sigma_{\vl}$ to the set of CV
$\rho_{\vk}$ and the approximation of the Fourier transform of the interaction potential
\be
\Phi(k)=\lb
\begin{array}{ll}
\Phi(0)(1-2b^2k^2), & k\leq B_0,\\
\Phi_0=\Phi(0)\bar\Phi, & B_0<k\leq B,
\end{array}
\rd \label{f3}
\ee
we start from the partition function in the following form:
\bea
Z&\propto&\int(d\rho)^{N_0}\exp[-a_1\sqrt N_0\rho_0-
\frac{1}{2}\sum_{k\leq B_0}d(k)\rho_{\vk}\rho_{-\vk}\non
& & -\frac{a_4}{4!N_0}\sum_{k\leq B_0}\rho_{\vk_1}...\rho_{\vk_4}\delta_{\vk_1+...+\vk_4}].
\label{f4}
\eea
Here $\bar\Phi$ is the small constant, $\delta_{\vk_1+...+\vk_4}$ is the
Kronecker symbol, $N_0=Ns_0^{-d}$, $B_0=B/s_0$. The quantity $B=\pi/c$ determines the
first Brillouin half-zone, $c$ is the lattice constant,
and the initial division parameter $s_0$ defines the region, where the parabolic approximation for
$\Phi(k)$ is valid [see Eq. (\ref{f3})]. The CV $\rho_{\vk}$ are associated with modes of spin moment density oscillations.
The variable $\rho_0$ corresponds to the system order parameter. In the general case, the expression
in the exponent of relation (\ref{f4}) is the $n$-power polynomial, where $n$
takes on the integer values. The polynomial with coefficients $a_n$ determines Jacobian of transition
to the set of CV \cite{prfref5,novo89}. Performing our calculations, we are restricted by the simplest
non-Gaussian distribution of the order parameter fluctuations with $n=4$ ($\rho^4$ model). In order to cancel
the cubic term, the corresponding substitution of variables is carried out.
The CV method allows us to perform a well controlled calculation at each order of the truncation. In this
respect, the term "model" may be associated with the order (models $\rho^4$, $\rho^6$, etc.).

The initial coefficients in (\ref{f4}) are expressed by relations
\bea
&&a_1=-s_0^{d/2}h',\qquad
d(k)=a_2+\beta\Phi_0-\beta\Phi(k), \non
&&a_2=1-3c_h^{''},\qquad
a_4=6c_h^{''}.
\label{f5}
\eea
Here $h'=\beta h$ is the dimensionless field, $\beta=1/kT$ is the inverse
temperature ($k$ is the Boltzmann constant), and $c_h^{''}=s_0^{-d}(1-4h'^2)/3$.
The first term in the exponent of the partition function
(\ref{f4}) is related to the presence of an external field. The coefficients  $a_2$ and $a_4$
depend on the field weakly.

The calculation of the partition function (\ref{f4}) is performed using the step-by-step integration
with respect to the CV. The procedure of such an integration is described in detail in \cite{prfref5,novo89,k05}.
For this purpose, we divide the phase space of the CV $\rho_{\vk}$ into layers with the division parameter $s$.
The integration begins from the variables with large values of the wave vector and terminates at
$\rho_{\vk}$ with $k\rightarrow0$. In each $n$th layer, which corresponds to the region of wave vectors
 $B_{n+1}<k\leq B_{n}$, where $B_{n+1}=B_n/s$, the Fourier transform of the potential $\Phi(k)$
is replaced by its average value. For simplicity of estimations, we do not take into account
the correction for the potential averaging. Such an approximation corresponds to the
LPA in the Wilson exact RG approach.
Inclusion of this correction leads to an appearance  of the critical exponent $\eta$,
which characterizes the behaviour of the pair correlation function for $T=T_c$.
Such calculations have been performed within CV method in the absence of an
external field. The formal part of the procedure has already been
presented in detail for the $\rho^4$ model in \cite{prfref5,novo89}. The quantity $\eta$ takes on the value $\eta\approx0.024$
\cite{p99}. For comparison, the exponent $\eta=0.044$ was obtained using the quartic effective potential
and method based on an exact flow equation for a coarse-grained free energy \cite{btw96}. Including the
above-mentioned correction and the related shift of the fixed point does not qualitatively change the main
thermodynamic characteristics of the system.

Thus, as a result of performing $n$ iterations, the partition function is presented in the form
\be
Z\propto Q_0Q_1...Q_nI_{n+1},
\label{f6}
\ee
where quantities
\be
Q_n=[Q(P^{(n-1)})Q(d_n)]^{N_n}
\label{f7}
\ee
are the partial partition functions of the  $n$th block structure. Here
\bea
Q(P^{(n-1)})=(2\pi P_2^{(n-1)})^{-\frac{1}{2}}\Bigl(1-\frac{3}{4}G^{(n-1)}\Bigl),\non
Q(d_n)=\Biggl(\frac{24}{a_4^{(n)}}\Biggr)^{\frac{1}{4}}\gamma_1\Bigl(1-\gamma h_2^{(n)}\Bigl)
\label{f8}
\eea
as well as
\bea
P_2^{(n-1)}=\Biggl(\frac{24}{a_4^{(n-1)}}\Biggr)^{\frac{1}{2}}\gamma\Bigl(1+t_2 h_2^{(n-1)}\Bigl),\non
G^{(n-1)}=s^{-d}G_0(1+G_2h_2^{(n-1)}),\non
h_2^{(n)}=\sqrt6 \frac{d_n(B_{n+1},B_n)}{(a_4^{(n)})^{{1/2}}}.
\label{f9}
\eea
The coefficients $\gamma$, $\gamma_1$, $G_0$, $t_2$ and $G_2$ are independent of the field and are given
in \cite{k05}. The quantity  $d_n(B_{n+1},B_n)=d_n(0)+qs^{-2n}$ is determined by the
averaged Fourier transform of the interaction potential $\Phi(B_{n+1},B_n)$
with the help of the relation
\be
d_n(k)=a_2^{(n)}+\beta\Phi(B_{n+1},B_n)-\beta\Phi(k).
\label{f10}
\ee
Here $q=\bar q\beta\Phi(0)$ and $\bar q=(b\pi/c)^2s_0^{-2}(1+s^{-2})$.
The quantity $I_{n+1}$ in Eq. (\ref{f6}) has the following form:
\bea
I_{n+1}&=&\int (d\rho)^{N_{n+1}}\exp\Biggl\{-a_1^{(n+1)}\sqrt N_{n+1}\rho_0\non
&& -\frac{1}{2}\sum_{k\leq B_{n+1}}d_{n+1}(k)\rho_{\vk}\rho_{-\vk}\non
&&-\frac{a_4^{(n+1)}}{4!N_{n+1}}\sum_{k\leq B_{n+1}}\rho_{\vk_1}...\rho_{\vk_4}\delta_{\vk_1+...+\vk_4}\Biggr\}.
\label{f11}
\eea
The coefficients $a_l^{(n+1)}$ are defined by recurrence relations (RR)
\bea
&&a_1^{(n+1)}=s^{d/2}a_1^{(n)},\non
&&a_2^{(n+1)}=f_{00}(a_4^{(n)})^{1/2}(1+\alpha_2h_2^{(n)}),\non
&&a_4^{(n+1)}=s^{-d}f_{01}a_4^{(n)}(1+\alpha_4h_2^{(n)}),
\label{f12}
\eea
where quantities $f_{00}$, $\alpha_2$, $f_{01}$  and $\alpha_4$ are  constants \cite{k05}.
Performing the scaling transformations $\rho_{\vk}=s\rho_{\vk}'$ with
introduction of the notations $\omega_n=s^na_1^{(n)}$,
$r_n=s^{2n}d_{n}(0)$, $u_n=s^{4n}a_4^{(n)}$ and taking into account the expression for $h_2^{(n)}$
from Eqs. (\ref{f9}), the RR (\ref{f12}) can be presented in the matrix form
\be
\\
\left( \begin{array}{c}
\omega_{n+1}-\omega^* \\
r_{n+1}-r^*\\
u_{n+1}-u^*
\end{array} \right) =
\left( \begin{array}{ccc}
R_{11} & 0 & 0 \\
0 & R_{22} & R_{24}\\
0 & R_{42} & R_{44}\end{array} \right)
\\
\left( \begin{array}{c}
\omega_n-\omega^* \\
r_n-r^*\\
u_n-u^*
\end{array} \right).
\label{f13}
\ee
Here $R_{11}=s^{{d+2}/{2}}$, $R_{22}=s^2f_{00}\alpha_2\sqrt6$, $R_{24}=s^2f_{00}u^{*-1/2}$,
$R_{42}=sf_{01}\alpha_4\sqrt6u^{*1/2}$ and $R_{44}=sf_{01}$.
The coordinates of the fixed point $w^*=0$, $r^*=-q$ and $u^*=q(1-s^{-2})/f_{00}$ are determined
using the RR.
The division parameter value $s=s^*=3.3783$ is defined by condition
\[
h_2^*=\sqrt{6} (r^*+q)(u^*)^{-1/2}=0.
\]
The fixed point corresponds to the presence of the long-range interactions with the infinite
correlation length. For describing the system behaviour, we should perform
the infinite number of iterations ($n\rightarrow\infty$). Nevertheless, the presence of the field
or the temperature value distinguishing from the critical one, leads to the deviation from the
fixed point  ($w^*,r^*,u^*$). In this case, the correlations become finite.
At small values of the field  $h$ ($h=0\div0.01$) and reduced temperature
$\tau=(T-T_c)/T_c$ ($\tau=0\div0.01$),  i.e., at small deviations from the fixed point, we
can use the linearized RR whose solutions are given by
\bea
& & \omega_n=\omega^*-s_0^{d/2}h'E_1^n,\non
& &r_n=r^*+c_{k1}^{(0)}\beta\Phi(0)\tau E_2^n+ c_{k2}T_{24}^{(0)}
(\varphi_0^{1/2}\beta\Phi(0))^{-1}E_4^n,\non
& & u_n=u^*+c_{k1}^{(0)}(\beta\Phi(0))^2T_{42}^{(0)}\varphi_0^{1/2}\tau E_2^n+c_{k2}E_4^n.
\label{f14}
\eea
Here $E_1=20.977$, $E_2=7.374$, $E_4=0.397$ are the eigenvalues of the RG linear transformation matrix,
$T_{24}^{(0)}=u^{*1/2}R_{24}(E_4-R_{22})^{-1}$,
$T_{42}^{(0)}=u^{*-1/2}R_{42}(E_2-R_{44})^{-1}$. The coefficients $c_{k1}^{(0)}$ and $c_{k2}$
are expressed by relations
\bea
c_{k1}^{(0)}&=&\Bigl[1-\bar\Phi-\bar q-T_{24}^{(0)}u_0\varphi_0^{-1/2}(\beta\Phi(0)\beta_c\Phi(0))^{-1}\non
&&-T_{24}^{(0)}\varphi_0^{1/2}\Bigr]\times\Bigl[1-T_{24}^{(0)}T_{42}^{(0)}\Bigr]^{-1},\non
c_{k2}&=&\Bigl[u_0-u^*-T_{42}^{(0)}\varphi_0^{1/2}\beta\Phi(0)(r_0-r^*)\Bigr]\non
&&\times\Bigl[1-T_{24}^{(0)}T_{42}^{(0)}\Bigr]^{-1},\no
\eea
where $\varphi_0=(\bar q(1-s^{*-2})/f_{00})^2$. These quantities do not depend on the field.
Using the  condition of the linearity of deviations from the fixed point, we can define the number of iterations
$n=n_p$. This quantity determines the size of the scaling region. Taking into account the values of
the eigenvalues $E_l$ and coefficients in terms with the variables  $h$ and $\tau$ in Eqs. (\ref{f14}),
one can see that the deviations from the fixed point are formed mainly in the
first two expressions for $\omega_n$ and $r_n$. Thus, the quantity  $n_p$
is found using the equation
\be
(-s_0^{d/2}h'E_1^{n_p+1})^2+(c_{k1}^{(0)}\tau \beta\Phi(0)E_2^{n_p+1})^2=r^{*2}.
\label{f15}
\ee
The right-hand side of Eq. (\ref{f15}) determines the maximal deviation, which satisfies  the linearity
condition. The choice of this quantity is also based on the condition related to the  coefficient $d_{n+1}(k)$.
When the block lattice size riches the correlation length, the coefficient $d_{n_p+1}(k)$ in expression
(\ref{f11}) in the case of $T>T_c$ changes sign and takes on the positive value. Therefore,
it is possible to perform the next step of integration of the partition function in the Gaussian approximation.

The quantity $n_p=n_p(\tau,h)$ (or the point of exit of the system from the scaling region) is a
function of field and temperature.
This characteristic provides the crossover between critical behaviour
controlled by temperature and one defined by the field variable. The equality of the left-hand side terms
in Eq. (\ref{f15}) corresponds to the case described in the field-temperature plane by relation
\be
\tau \propto h^{1/\beta\delta}.
\label{f16}
\ee
In the vicinity of values of the field and temperature satisfying relation (\ref{f16}),
the scaling variable is of the order of unity and the series expansions in the scaling variable are not effective.
It should be noted, that the RR  (\ref{f13}) and their solutions  (\ref{f14}) are valid in the
high-temperature region as well as in the case of  $T<T_c$ at $n\leq n_p$.

\section{Critical exponents and thermodynamic characteristics of the system}

\subsection{The case of $T>T_c$}

As was mentioned above, the block lattice size after the $n_p$ iterations is com\-men\-su\-rate
with the correlation length. Increasing the number of iterations  $n$ and excluding
"non-essential" variables $\rho_{\vk}$, we renormalize the coefficients $a_l^{(n)}$.
At $n>n_p$, due to changing the sign of the coefficient in the quadratic term in the exponent of
relation (\ref{f11}), we can calculate the quantity  $I_{n+1}$ appearing in (\ref{f6})
on the basis of the Gaussian approximation. In the case of  $T>T_c$, the coefficient
$d_{n_p+1}$ increases gradually and, for the quantitative results, one should
perform the additional step of integration \cite{kpprevb06}.
Then the partition function (\ref{f6}) takes on the form
\be
Z\propto Q_0Q_1...Q_{n_p+1}I_{n_p+2},
\label{f17}
\ee
where the quantity $I_{n_p+2}$ has the form similar to expression (\ref{f11}).
After per\-for\-ming this iteration, the quadratic term in  $I_{n_p+2}$
dominates over other terms for all $k\neq0$.
Calculating the integral $I_{n_p+2}$ with respect to variables  $\rho_{\vk}$ with indices $0<k\leq B_{n_p+2}$,
we obtain
\be
I_{n_p+2}=\prod_{k\leq B_{n_p+2}\atop{k\neq0}}\left(
\frac{\pi}{d_{n_p+2}(k)}\right) ^{\frac{1}{2}}I^{(0)}_{n_p+2}.
\label{f18}
\ee
The quantity
\bea
I_{n_p+2}^{(0)}&=&\int
(d\rho_0)\exp \Biggl[N^{1/2}h'\rho_0-\frac{1}{2}d_{n_p+2}(0)\rho^2_0\non
& &-\frac{1}{4!}a_4^{(n_p+2)} N^{-1}_{n_p+2}\rho^4_0\Biggr]
\label{f19}
\eea
is crucial for system description by two reasons.  Firstly,
as a result of the RG transformations, every next block lattice includes the previous one
and exhibits its physical properties. Since, the integration with respect to the variable $\rho_0$
is the last step of calculation, this contribution is the most essential for the system description.
The case of  $k=0$ corresponds to the total renormalization of the coefficients of the
distribution of the order parameter fluctuations. Secondly, the average value of the variable $\rho_0$
is the order parameter of the system \cite{prfref5,novo89}.
Thus, performing the substitution of the variable $\rho_0=\sqrt{N}\bar\rho_0$ and
evaluating the integral  (\ref{f19}) by the steepest-descent method, we obtain the
expression for the system free energy
\bea
F_{\vk=0}^{(+)}&=&-NkT s^{-3(n_p+1)} \Bigl[h'\sigma_0 s^{\frac{5}{2}(n_p+1)}-\frac{1}{2}r_{n_p+2}s^{-2}\sigma^2_0 \non
&&-\frac{1}{4!}s_0^{d}s^{-1}u_{n_p+2}\sigma^4_0\Bigr].
\label{f20}
\eea
The quantity $\sigma_0$ is the root of the cubic equation
\be
h' s^{\frac{5}{2}(n_p+1)}-r_{n_p+2}s^{-2}\sigma_0-\frac{1}{6}s_0^{d}s^{-1}u_{n_p+2}\sigma^3_0=0,
\label{f201}
\ee
which is obtained using the extremum condition for the expression in the exponent
of the relation (\ref{f19}) with respect to the variable
$\rho_0$. Here, we use the substitution of variable
\be
\bar\rho_0^{(+)}=\sigma_0s^{-\frac{n_p+1}{2}}.
\label{f21}
\ee
The choice of the root of Eq. (\ref{f201}) is based on the condition of the minimum for the free energy.
The contributions related to variables  $\rho_{\vk}$ with nonzero values of the wave vector are small
in comparison with the contribution (\ref{f20}). The integration with respect to these variables is
necessary for obtaining the renormalization coefficients appearing in expression  (\ref{f20}),
which is the system free energy at $T>T_c$. The relation  (\ref{f21})  corresponds to the  equation of state.

\subsection{The case of $T<T_c$}
Performing the integration of expression  (\ref{f11}) in the case of the low-temperature region,
the displacement procedure proposed by Yuhnovskii \cite{novo89}  is used.
It is pro\-vi\-ded by the substitution of the variables
\be
\rho_{\vk}=\eta_{\vk}+\sigma_h\sqrt{N}\delta_{\vk}.
\label{f22}
\ee
After performing such a substitution, the coefficient in the quadratic term in Eq. (\ref{f11})
becomes positive and dominating, as in the case of $T>T_c$. It takes on the form
\be
d_h(k)=d_{n_p+1}(k)+\frac{1}{2}a_4^{(n_p+1)}\sigma_h^2\frac{N}{N_{n_p+1}}.
\label{f23}
\ee
The detail procedure for obtaining this expression is given in \cite{mycmp05}.
Calculating the integral  $I_{n_p+1}$ with respect to the variables  $\eta_{\vk}$ at $k\neq0$
in the Gaussian approximation and returning to the variable $\rho_0$, we obtain
\be
I_{n_p+1}=\prod_{k\leq B_{n_p+1}\atop{k\neq0}}\left(
\frac{\pi}{d_{h}(k)}\right) ^{\frac{1}{2}}I^{(0)}_{n_p+1},
\label{f24}
\ee
where
\bea
I_{n_p+1}^{(0)}&=&\int
(d\rho_0)\exp \Biggl[N^{1/2}h'\rho_0-\frac{1}{2}d_{n_p+1}(0)\rho^2_0\non
& &-\frac{1}{4!}a_4^{(n_p+1)} N^{-1}_{n_p+1}\rho^4_0\Biggr].
\label{f25}
\eea
Performing the integration in expression for $I_{n_p+1}^{(0)}$ by the same method as in the case of  $T>T_c$,
we find the system free energy
\bea
F_{\vk=0}^{(-)}&=&-NkT s^{-3(n_p+1)} \Bigl[h'\sigma_h s^{\frac{5}{2}(n_p+1)}-\frac{1}{2}r_{n_p+1}\sigma_h^2 \non
&&-\frac{1}{4!}s_0^{d}u_{n_p+1}\sigma^4_h\Bigr]
\label{f26}
\eea
and equation of state
\be
\bar\rho_0^{(-)}=\sigma_h s^{-\frac{n_p+1}{2}}
\label{f27}
\ee
in the low-temperature region.

\subsection{Critical exponents and scaling functions}
The free energies  (\ref{f20}), (\ref{f26}) and equations of state (\ref{f21}),  (\ref{f27})
in the cases of  $T>T_c$ and $T<T_c$ are explicit analytic expressions with coefficients,
which depend on the microscopic parameters, in particular, on the lattice constant and
parameters of the interaction potential. At $h=0$, taking into account Eqs. (\ref{f15}) and (\ref{f27}),
we get the scaling representation
\be
\bar \rho_0^{(-)}=B' \tau^{\beta}, \qquad T<T_c,
\label{f28}
\ee
where $\beta=\nu/2$, $B'=\sigma_h (c_{k1}^{(0)}/\bar q)^{\beta}$, and the critical exponent
of the correlation length $\nu$ is calculated by the formula
\be
\nu=\frac{\ln{s}}{\ln{E_2}}\approx0.609.
\ee
The value of the critical exponent $\nu=0.609$ corresponds to the $\rho^4$ model app\-ro\-xi\-ma\-tion with
$\eta=0$. In \cite{btw96}, the quantity $\nu=0.643$ is found for the similar approximation
(the quartic potential but $\eta\neq0$). As was shown in \cite{prfref7,ypk02,ypk022}, the better convergent results
within the approximation $\eta=0$ can be obtained by using more complicated models. In particular,
$\nu=0.637$, $\nu=0.646$ and $\nu=0.649$ for models $\rho^6$, $\rho^8$ and $\rho^{10}$, respectively \cite{prfref7}.
The critical exponent of the correlation length for O(1) model calculated using Wegner-Houghton RG equation takes
on the values $0.527$, $0.586$, $0.673$ and $0.759$  for the second, third, fourth and fifth order
field truncations (corresponding to $\phi^4$, $\phi^6$, $\phi^8$ and $\phi^{10}$), respectively \cite{mop88}.

\begin{figure}[htb]
\centerline{\includegraphics[width=0.55\textwidth]{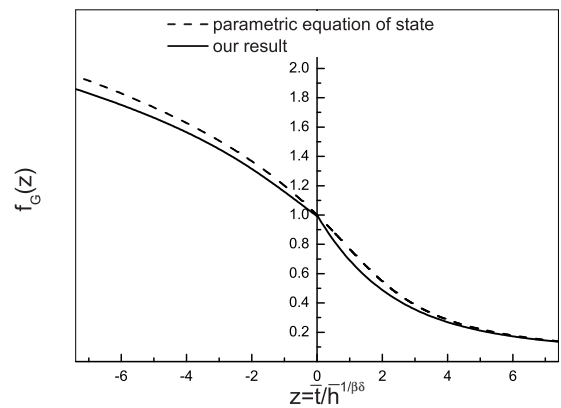}}
\caption{The scaling function of the order parameter. The solid line corresponds to Eqs. (\ref{f21}) and (\ref{f27})
at $T>T_c$ and $T<T_c$, respectively. The dashed line corresponds to the parametric representation of the equation of
state \cite{gz197} and MC simulations  \cite{efs03}.} \label{fff1}
\end{figure}

On the critical line  $\tau=0$ at $h>0$, we have
\be
\bar \rho_0^{(\pm)}=D_c^{-1/\delta} h'^{\frac{1}{\delta}},
\label{f29}
\ee
where  $D_c=r^*/(\sigma_0^{\delta}s_0^{d/2})$, and the critical exponent
of the order parameter $\delta$ is defined as
\be
{\delta}=\frac{2\ln E_1}{\ln s}={5}.
\label{f30}
\ee
Note that quantities $\sigma_0$ in Eq. (\ref{f21}) and $\sigma_h$ in Eq. (\ref{f27})
coincide on the critical line.
In order to compare the results with others, we use the explicit form for the scaling functions.
In Fig. \ref{fff1}, the equations of state  (\ref{f21}) and (\ref{f27}) are presented
in the scaling form
\be
f_G(z)=M\bar h^{-1/\delta}
\ee
(the solid line), where $z=\bar t/\bar h^{1/\beta\delta}$ is the scaling variable. Here $\bar h=h'/D_c$, $M\equiv\bar \rho_0$, and
$\bar t=\tau B'^{1/\beta}$. Our results accord qualitatively with the well-known parametric representation of the
equation of state \cite{gz197} and Monte-Carlo (MC) results \cite{efs03} (the dashed line). It should be noted
that the results obtained in this work do not require any parametrization and allow one to take into account
the influence of the microscopic parameters of the Hamiltonian (\ref{f1})on the critical amplitudes.

In Fig. \ref{fff2}, the system susceptibility is given in the scaling form
\be
f_{\chi}(z)=\chi D_c\bar h^{1-1/\delta},
\ee
where $\chi=\partial M/\partial h'$.

Some disagreement of our data with ones obtained in \cite{gz197} can be connected with the
model approximation, since present calculations are carried out using the simplest non-Gaussian model $\rho^4$.

\section{Conclusions}
\begin{figure}[htb]
\centerline{\includegraphics[width=0.55\textwidth]{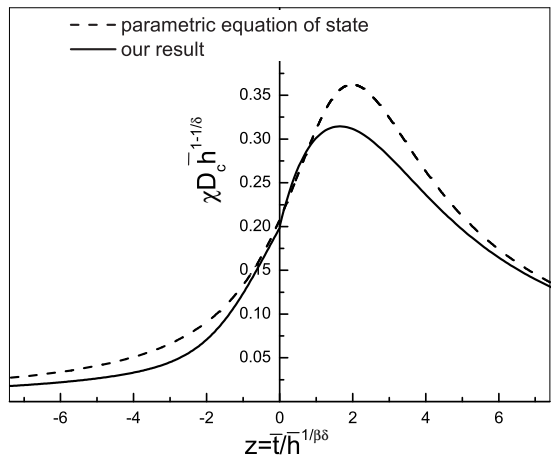}}
\caption{The scaling function of the susceptibility (the solid line is the CV method result,
the dashed line is the result obtained in \cite{gz197,efs03}).} \label{fff2}
\end{figure}

The description of the $3D$ Ising-like magnet behaviour in the vicinity of the critical point is proposed.
The calculations are carried out by the CV method with using the RG transformations of the Wilson type.
It is established that contributions related to nonzero values of the wave vector are small in comparison with
one from the variable  $\rho_0$, the average value of which is the system order parameter.
The dependence of the limit magnitude of the long-range interactions on the temperature and field variables is presented.
The scaling region size is determined as function of temperature and field by using Eq. (\ref{f15}).
The universal and nonuniversal characteristics of the system are obtained simultaneously.
Particularly, the analytic expressions for the free energy and equation of state, which are valid in the critical region
in the whole field-temperature plane, are presented.
The system susceptibility is constructed in the scaling form.
As is seen from Fig. \ref{fff2}, the susceptibility maximum location on the temperature scale is in good qualitative
agreement with one obtained using the parametric representation of the equation of state and MC simulations.

The calculations of system nonuniversal characteristics are based on the physically well-grounded and
mathematically rigorous use of the phase space of CV. They are per\-for\-med starting from the Hamiltonian,
which contains the microscopic pa\-ra\-me\-ters. The results are obtained using the quartic approximation
for the effective Hamiltonian ($\rho^4$ model) without taking into account the  correction for the potential averaging
(analog of the LPA). Due to removing the cubic term by making the shift of CV, the presence of the field is represented
only by the  linear term. This result is valid also for models $\rho^6$, $\rho^8$, etc. In other words,
the presence of the field caused to appearance of the additional relation for $\omega_n$ in Eq. (\ref{f14}).

The similar estimations can be generalized to the  models with an $n$-component order parameter


\end{document}